# Two-octave-wide (3–12 μm) mid-infrared frequency comb produced as an optical subharmonic in a nondispersive cavity

Q. Ru[1], T. Kawamori[1], P. G. Schunemann[2], S. Vasilyev[3], S. B. Mirov[3,4], K. L. Vodopyanov[1*]

[1] CREOL, College of Optics and Photonics, Univ. Cent. Florida, Orlando, FL 32816, USA
[2] BAE Systems, P. O. Box 868, MER15-1813, Nashua, New Hampshire 03061-0868, USA
[3] IPG Photonics-Mid-Infrared Lasers, Birmingham, Alabama 35203, USA
[4] Dept. of Physics, Univ. of Alabama at Birmingham, Birmingham, Alabama 35294, USA
* vodopyanov@creol.ucf.edu

**Coherent laser beams in the 3 to 20 μm region of the spectrum are most applicable for chemical sensing by addressing the strongest vibrational absorption resonances of the media. Broadband frequency combs in this spectral range are of special interest since they can be used as a powerful tool for molecular spectroscopy offering dramatic gains in speed, sensitivity, precision, and massive parallelism of data collection. Here we show that a frequency comb realized through subharmonic generation in an optical parametric oscillator (OPO) based on orientation-patterned gallium phosphide (OP-GaP) pumped by a Kerr-lens mode-locked 2.35-μm laser can reach a continuous wavelength span of 3–12 μm, thus covering most of the molecular 'signature' region. The key to achieving such a broad spectrum is to use a low-dispersion cavity entailing all gold-coated mirrors, minimally dispersive and optically thin intracavity elements, and a specially designed pump injector. The system features a smooth ultrabroadband spectral output that is phase coherent to the pump laser comb, 245 mW output power with high (>20%) optical conversion efficiency, and a possibility to reach close to unity conversion from a mode-locked drive, thanks to the non-dissipative downconversion processes and photon recycling.**

## Introduction

Mid-infrared spectroscopy provides rich and quantitative information on the the structure of matter in the gas, liquid, and solid phases, offering applications ranging from trace molecular sensing and environmental monitoring to the study of combustion dynamics, medical exhaled breath analysis, and nano-IR spectroscopy. Optical frequency combs in the mid-infrared (mid-IR) region are powerful photonic tools that can be used in the most advanced spectroscopic techniques, such as multi-heterodyne dual-comb spectroscopy enabling rapid, broadband and precise spectral analysis [1,2].

A variety of approaches have been developed in the last decade to generate frequency combs across the mid-IR that include mode-locked Tm-doped fiber [3] and Cr:ZnS solid-state lasers [4], comb sources based on difference-frequency generation (DFG) [5-15], intrapulse DFG (IDFG) [16-25], optical parametric oscillators (OPOs) [26-28], supercontinuum (SC) generation in fibers and waveguides [29-34], quantum cascade lasers (QCL) [35,36], interband cascade lasers (ICL) [37], microresonators (μ-res) [38-42], and electro-optic modulation (EOM) [43].

For multi-species spectroscopic detection, e.g. in medical breath analysis targeting a big variety of biomarkers simultaneously, one needs a spectrally broad comb with relatively flat spectrum and enough power to achieve a reasonable signal-to-noise ratio. Fulfilling these metrics simultaneously has been challenging. For example, mode-locked lasers operate only in the short-wavelength portion (λ<3 μm) of the mid-IR; the other techniques suffer from either low output power (μ-res), limited frequency span (DFG, OPO, QCL, ICL, EOM combs), strongly uneven spectrum (SC, QCL, μ-res), or difficulty in referencing the optical frequency to the primary standards (QCL, ICL). There has been a number of recent reports on IDFG sources that combine Watt-level output with multi-octave instantaneous spectra in the mid-IR [20, 21, 22, 24]. However, the output of those ultra-broadband sources does not constitute a single comb – rather it consists of several superimposed combs with different carrier–envelope offsets and most power concentrated near the wavelength of the driving laser. A diagram in Fig. 1 illustrates the most advanced, in terms of spectral span and the average power, comb generators across the mid-IR.

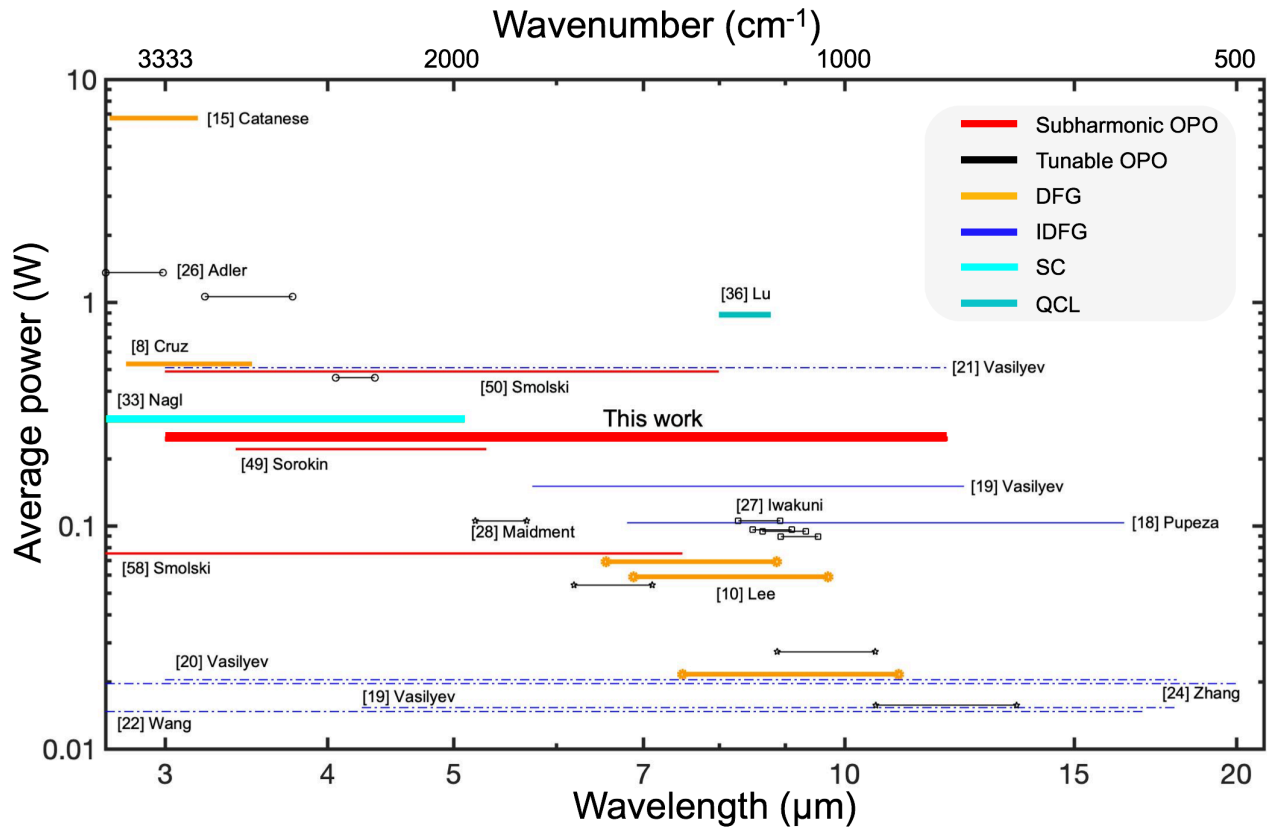

**Fig. 1. Frequency span and average power of selected frequency comb sources across the mid-IR**. Dash-dotted lines represent IDFG spectra [20-24] that are composed of overlapping bands with different carrier–envelope offsets.

Subharmonic optical parametric oscillators (OPOs) are noteworthy sources of mid-IR frequency combs [44-50], now widely used in spectroscopic studies [51,52], random number generators [53], coherent Ising machines [54], and in the study of solitons [55]. It is an ideal coherent frequency divider without any excess phase noise, which rigorously both down-converts and augments the spectrum of a near-IR pump frequency comb [56-59]. Its main benefits are low oscillation threshold, broad bandwidth, and excellent stability when actively locked for the doubly-resonant operation at degeneracy. It's another advantage is high conversion efficiency from the pump laser that has been shown to exceed 50% [60].

Here we report on a new approach to producing an optical subharmonic that allows to achieve by far the broadest ever obtained, in excess of 75 THz, fully coherent optical frequency comb in the mid-IR at $\lambda>3$ μm. The comb consists of 950,000 spectral components that share common offset and smoothly cover the mid-IR range from 3 to 12 μm. This was enabled by an unconventional long-wave pump, a minimally dispersive resonant cavity utilizing only metallic mirrors, and a way the pump beam is injected into the cavity. This design has allowed achieving a regime where the three-wave parametric processes are enhanced by the four-wave interactions. We also developed a theoretical model that provides an insight into this new approach to generate multi-octave frequency combs with highly coherent smooth spectra, Watt-level output power, and conversion efficiency that can, in theory, approach unity.

## Setup

The pump source (Fig. 2a) was a Kerr-lens mode-locked $Cr^{2+}$:ZnS laser with 2.35-μm central wavelength, 1.2-W average power, 79-MHz repetition frequency, and bandwidth-limited pulse duration of 62 fs [19]. The bowtie ring OPO cavity was composed of gold-coated mirrors only. Two of these mirrors (M2 and M3) were parabolic with an off-axis angle of 30° and 30-mm radius of curvature in the apex (effective focal distance 16 mm). The other two mirrors were flat (not shown are additional two pairs of folding mirrors used to reduce the footprint). To incouple the laser pump beam, we used a specially designed pump injector – a dielectric mirror inside the OPO cavity on a thin (0.5-mm) wedged (0.5° angle) ZnSe substrate. The coating on one side was such that the injector had high reflection (>90%) for the 2.35-μm pump and high transmission for the OPO signal/idler waves (>90% at 3.5-7 μm and >50% at 2.8-12 μm, Fig. 3a). The second face of the injector mirror was antireflection coated for 3-12 μm. The injector introduced a minimal group delay dispersion (GDD) in the OPO cavity – mostly coming from the ZnSe substrate whose group velocity dispersion (GVD) is very small within the OPO span (zero GVD crossing of ZnSe is at 4.81 μm). The injector approach allowed using all gold-coated mirrors, thus minimizing the cavity loss over the whole OPO wavelength span. Another element, an uncoated ZnSe wedge (0.3-0.8 mm thick, 1° angle) was placed inside the cavity to (i) fine tune the net intracavity GDD and (ii) vary the OPO cavity outcoupling via Fresnel reflection. A broad-bandwidth parametric gain was provided by a thin, 0.5-mm-long, orientation-patterned gallium phosphide (OP-GaP) – a newly developed quasi-phase matched (QPM) nonlinear crystal with an indirect bandgap of 2.26 eV, second-order nonlinear coefficient $d_{14}$=35 pm/V [61], and wide transparency range of 0.55–13 μm [28]. The crystal had a QPM reversal period of Λ=110 μm, designed for subharmonic generation from the 2.35-μm pump, and was placed at the Brewster's angle (71°). Inside the crystal, the beams propagated along $<\bar{1}10>$ and were co-polarized along <111> crystallographic direction. With its zero GVD crossing at λ≈4.8 μm, combined with small length, the OP-GaP provided a two-octaves-wide gain bandwidth and introduced minimum GDD. Fig.3a plots the injector transmission and the normalized OP-GaP parametric gain vs. wavelength, while Fig. 3b shows a computed extra phase accumulated per cavity roundtrip due to high-order (≥3) dispersion in the cavity; also shown are contributions from individual elements.

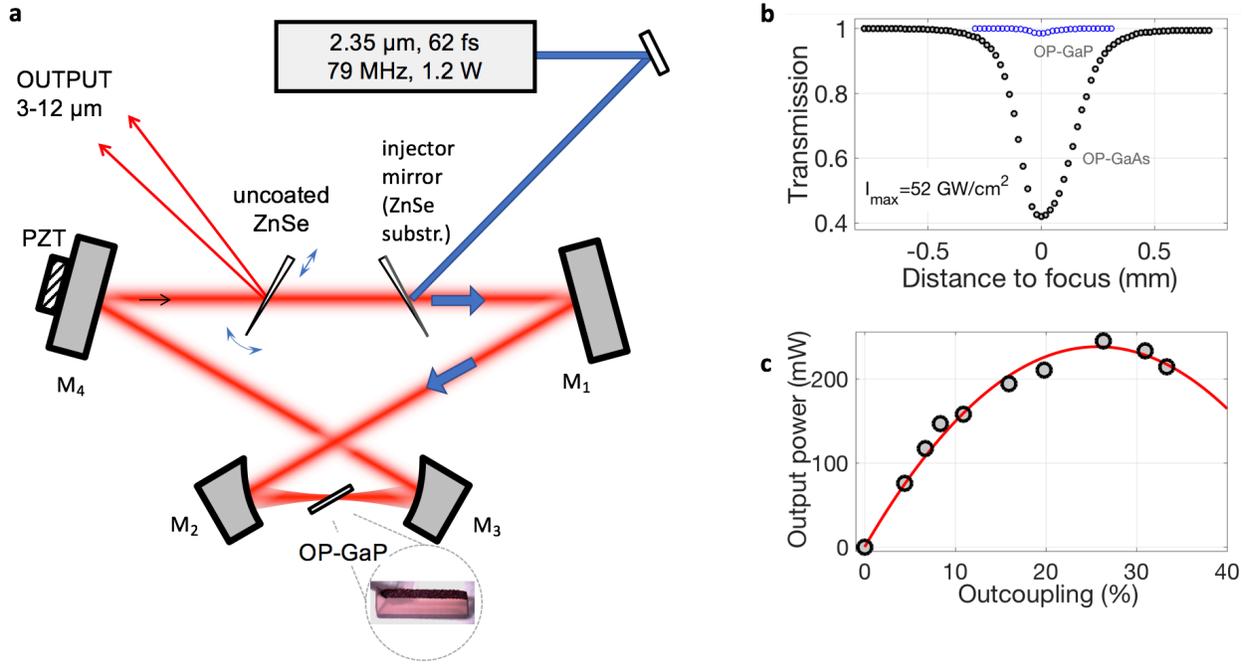

**Fig. 2. Subharmonic OPO schematic. a**, The ring-cavity OPO is synchronously pumped by a mode-locked Cr:ZnS oscillator. M2, M3, parabolic gold-coated mirrors. M1, M4, flat gold-coated mirrors. PZT, piezoelectric transducer for tuning the cavity length. **b**, Transmission of the focused 2.35-µm pump radiation as a function of the distance of a sample from the focus: a comparison between OP-GaP and OP-GaAs crystals. **c**, The OPO output power versus outcoupling strength. The solid curve is a trace for the eye.

To evaluate the role played by multi-photon absorption (MPA) at our level of pump intensities, we performed simple open-aperture z-scan measurement and compared the nonlinear absorption for the OP-GaP with that of another advanced mid-IR nonlinear material, an orientation-patterned gallium arsenide (OP-GaAs, bandgap 1.42 eV, $d_{14}$=94 pm/V). Fig. 2b shows variation of the transmission as the samples (both 0.5-mm-long, placed at the Brewster's angle) were translated through a 2.35-µm beam focus, with the peak intensity inside crystals reaching 52 GW/cm$^2$. While, for the OP-GaAs we observed a 60% transmission dip due to 3-photon absorption (3PA), in good accordance with its bandgap and 3PA coefficient of 0.35 cm$^3$/GW$^2$ [62, 63], we measured only a 1.5% dip for the OP-GaP. More detailed MPA studies (to be published elsewhere) indicate that the dominant nonlinear loss mechanism in the OP-GaP at 2.35 µm is 5-photon absorption, with an estimated 5PA coefficient of ~2.6×10$^{-6}$ cm$^7$/GW$^4$. The key result here is that at our level of pumping MPA plays a minor role in gallium phosphide.

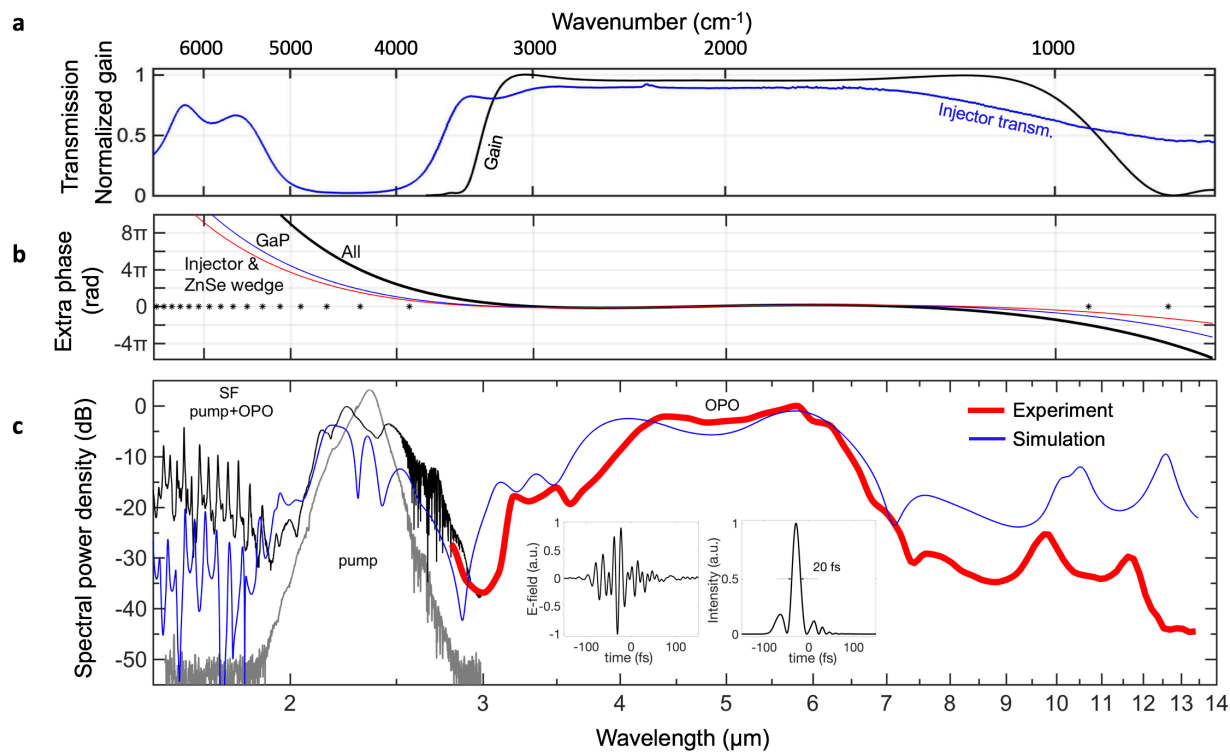

**Fig. 3. The OPO cavity characteristics and the spectrum**. **a**, Injector transmission and normalized parametric gain vs. wavelength. **b**, Computed extra phase per cavity roundtrip (black) and contribution of separate elements: OP-GaP (blue) and injector plus ZnSe tuning wedge (red). The asterisks indicate the spectral points where the extra phase acquires multiples of $2\pi$. **c**, A combined spectrum including the OPO spectrum (red), the original (incoming) spectrum of the pump (gray), and modified pump spectrum plus sum frequency (black). The simulated overall spectrum is shown in blue. The ripples near 2.7 µm are due to water absorption in the atmosphere. Inset: simulated time-domain profiles for the *E*-field and intensity of the OPO pulse (see Methods).

## Results

### Sub-harmonic comb spectrum

Since the OPO is doubly resonant, it only oscillates within discrete bands of the cavity length. Each band has a width that is a fraction of the pump wavelength, and adjacent bands are separated by approximately the pump wavelength (2.35 µm) in terms of the roundtrip cavity-length [56]. Fig. 4a shows the measured spectrum as a function of the cavity-length detuning. The OPO operated in a mode, where a piezo actuator attached to one of the mirrors was driven with a linear ramp, tuning the OPO length sequentially through several resonances over ~ 20 µm. The spectrum was measured with a grating monochromator and a mercury cadmium telluride (MCT) detector operating at 77 K, with 12-µm cut-off. Fig. 4b presents a simulation (see Methods) that included all possible three- and four-wave interactions inside the OP-GaP crystal, while Fig. 4c shows a simulation that took into account only three-wave interactions.

A continuous OPO operation was achieved by locking the cavity length (via dither-and-lock technique [45]) to a resonance that produced the broadest spectrum (second stripe from top in Fig. 4a). By further tuning the cavity dispersion via adjusting the thickness of the ZnSe wedge, a two-octave wide spectrum, 3-12 µm (at -37 dB level), was achieved (Fig. 3c).

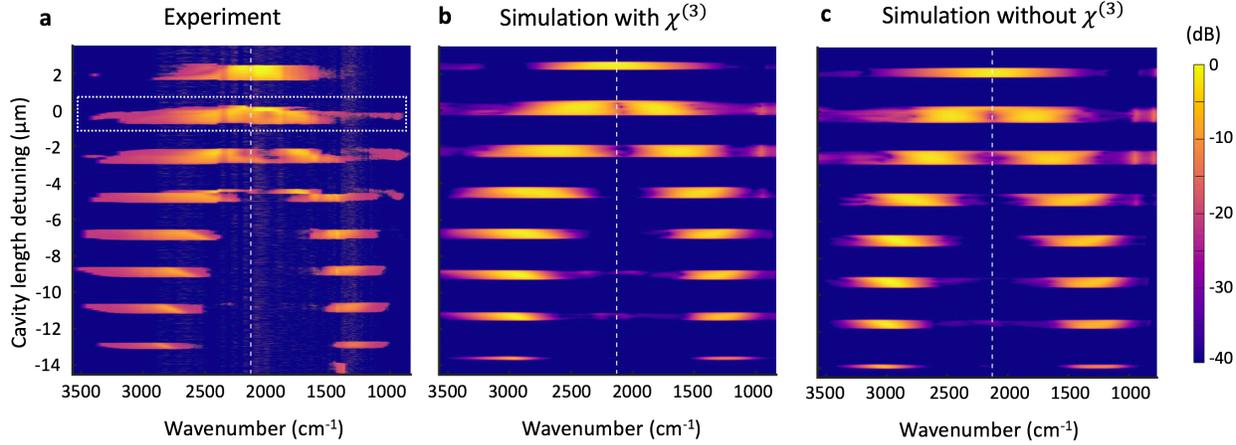

**Fig. 4. Color-coded intensity spectrum vs. cavity length detuning.** **a**, Experimental spectrum **b**, Simulated spectrum with both $\chi^{(2)}$ and $\chi^{(3)}$ included. **c**, Simulated spectrum with only $\chi^{(2)}$ included. The top stripes correspond the subharmonic OPO regime, while the lower ones correspond to the non-degenerate regime with distinct signal and idler bands.

To assess the spectrum at shorter waves, the OPO output was coupled to a single-mode fiber connected to optical spectrum analysers (Yokogawa AQ6376 for the 1.5–3 µm range; Agilent 86142B for the 1–1.5 µm range) with the resolution set to 2 nm. One can see that the spectrum of the pump laser (Fig. 3c, black curve) is broadened with respect to the initial, incoming, spectrum (gray curve) – a result of the frequency back conversion from the OPO. The spectrum at λ< 2 µm was produced via parasitic sum-frequency (SF) generation in the OP-GaP, between the pump and the OPO waves. The oscillations in the SF spectrum at 1.5 –2 µm are due to the fact that the OPO cavity is partially resonant here (Fig. 3a). However, due to large cavity GDD in this range, the roundtrip phase experiences rapid variations with wavelength (Fig. 3b), with peaks observed when the phase is a multiple of $2\pi$ (as indicated by the asterisks in Fig. 3b). For the same reason, the spectral peaks appear in the longwave portion of the spectrum, at 9.8 µm and 11.8 µm, – where the accumulated extra phase reaches $2\pi$ and $4\pi$ respectively. The simulated spectrum shown in Fig. 3c (blue curve) accounts well for the observed phenomena – both in the long-wave and short-wave portions of the spectrum

It appears that, in addition to $\chi^{(2)}$, the cubic nonlinearity $\chi^{(3)}$ in the OP-GaP importantly contributes to the OPO spectral broadening. In fact, because of the high value of nonlinear refraction for GaP (an estimated mid-IR value $n_2=1.9 \times 10^{-18}$ m$^2$/W, see Methods) and high circulating peak OPO intensity inside the OP-GaP crystal, of the order of $I_{max} \sim 100$ GW/cm$^2$, the dynamic phase shift induced by self-phase modulation (SPM) can reach the value of $\Delta \varphi = \frac{2\pi}{\lambda} n_2 I_{max} l = 1.3 \, rad$ (here λ is the center OPO wavelength, and $l$ is the length of the crystal). This is consistent with our experimental observations; for example, when the OPO is operating in a non-degenerate mode (bottom of Fig. 4a), one can see that the horizontal stripes corresponding to the signal wave on the blue side of the spectrum are broader than those for the idler wave. This is contrary to what might be expected from the 3-photon process: the widths of the stripes (in frequency units) should be identical because the photons are created in pairs with their frequencies equally spaced from the degeneracy point. The asymmetry can be attributed to the SPM-induced spectral broadening; the latter is proportional to the midpoint frequency and thus is larger for the signal wave. The simulated spectra of Fig. 4(b,c) support this idea – one can see that the asymmetry of the non-degenerate spectrum appears only when $\chi^{(3)}$ effects are turned on (Fig. 4b). Yet another $\chi^{(3)}$ effect – four wave mixing (FWM) – contributes to the expanding and smoothening of the long-wave portion of the OPO spectrum. In fact, for the degenerate FWM process $2\omega_1 = \omega_3 + \omega_4$, the phase matching condition $2k_1 = k_3 + k_4 + 2\gamma P$ [68] can be satisfied in the OP-GaP, thanks to its weak anomalous dispersion at >4.8 μm (here $\omega_i$ are angular frequencies, $k_i$ are wavevector modules, $\gamma = n_2\omega/cA_{eff}$ is the nonlinear FWM parameter, $c$ is the speed of light, $A_{eff}$ is an effective beam area, and $P$ is the peak power of the beam at $\omega_1$). For example, an energy transfer from the 6-μm part of the OPO spectrum to the weaker 11-μm portion through the above FWM process with $\omega_1 = 2\pi c/(6\mu m)$, $\omega_3 = 2\pi c/(11\mu m)$, and $\omega_4 = 2\pi c/(4.5\mu m)$ satisfies the phase matching condition. These effects are automatically included in our simulations (see Methods).

To optimize the OPO output power, we measured it as a function of the outcoupling strength by varying the angle of the ZnSe wedge (Fig. 2a). The average output power reached 245 mW (Fig. 2c) in two Fresnel reflections from the wedge at 26% outcoupling (combined from its two surfaces). The OPO pump threshold was 55 mW for 3.4% outcoupling and 250 mW for 26% outcoupling. A typical pump depletion was as high as 83%.

**Validation of the comb coherence**

To verify that the subharmonic OPO output is a single coherent comb, we performed beatnote measurements in the radiofrequency (RF) domain. Fig. 5a shows a combined spectrum – that of the OPO, the pump laser, SF, and second harmonic (SH) of the pump. First, we observed *f*-to-*2f* beat notes within the OPO spectral span, in the region of the overlap between the second harmonic of the 7-μm portion, produced parasitically in the OP-GaP crystal, and the OPO fundamental frequency at 3.5 μm (Fig. 5a), by using an InSb (77K, 60 MHz) photodetector and a bandpass (3.5±0.25 μm) filter. The beat note signal recorded with an RF spectrum analyzer showed the beat note 30 dB above the noise floor. It straightforwardly reveals the carrier envelope offset (CEO) frequency of the OPO, since $f_{beat} = 2f_{ceo}^{OPO} - f_{ceo}^{OPO} = f_{ceo}^{OPO}$ (Fig. 5b). The width of the beat note peak (~3 MHz at -3 dB) was somewhat broader than typically observed in such measurements. This was the result of the fluctuating offset frequency of the pump laser caused in turn by a modulated output of a multi-longitudinal-mode fiber laser used to pump the Cr:ZnS crystal. The beat note frequency was not random but rather dependent on the Cr:ZnS laser regime, e.g. on air humidity inside the laser cavity. As an additional proof that the beat notes are related to the CEO frequency of the OPO, we repeated the beat measurements at a neighboring OPO cavity-length resonance. Since the subharmonic OPO acts as a coherent frequency divider [57], the CEO frequency for the adjacent cavity-length resonances toggles between the two values [56]:

$$f_{ceo}^{OPO} = \frac{1}{2}f_0 \text{ and } f_{ceo}^{OPO} = \frac{1}{2}f_0 \pm \frac{1}{2}f_r, \quad (1)$$

where $f_0$ is the CEO of the pump laser and $f_r$ is the pulse repetition frequency. The beat notes corresponding to these two scenarios are shown in Fig. 5b and obey Eq. (1), namely the observed RF peaks are shifted by $\frac{1}{2}f_r$ with respect to each other.

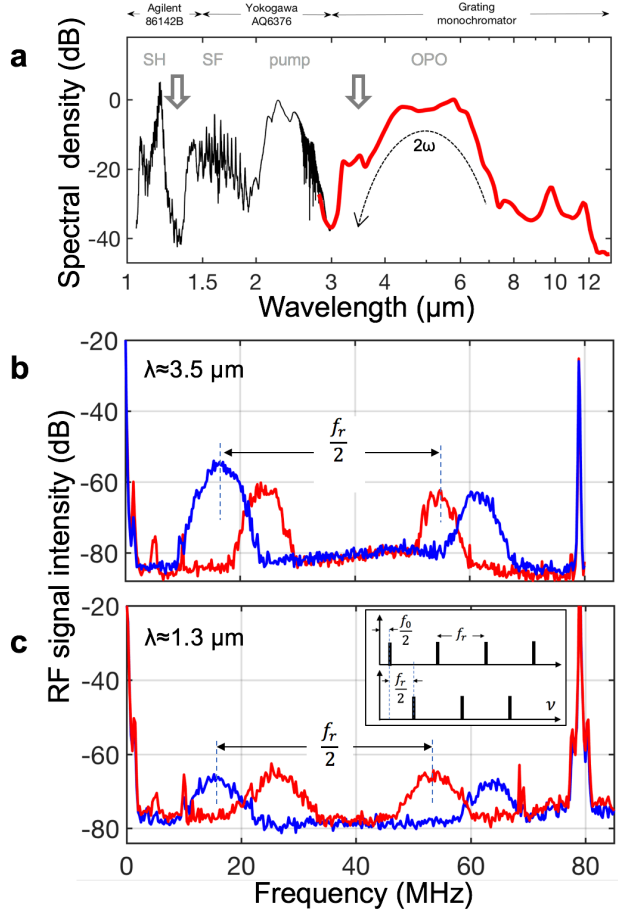

Fig. 5. Detection of RF beatnotes. **a**, Near-IR to mid-IR spectrum, taken with optical spectrum analyzers (Agilent 86142B and Yokogawa AQ6376) and a grating monochromator, which includes the spectrum of the OPO, of the pump, of the sum frequency (SF), and of the second harmonic (SH) of the pump. The arrows indicate the regions where RF beatnotes were recorded. **b**, RF spectrum showing *f*-to-2*f* beatnotes between the 3.5-μm and frequency doubled 7-μm portions of the OPO spectrum. **c**, beatnotes between the overlapping SH and SF at 1.3 μm. The blue and red curves in (b,c) correspond to the two neighboring cavity-length resonances. The inset shows the OPO modes extrapolated to zero frequency for the two adjacent cavity resonances.

Similarly, we used a spectral portion around 1.3 μm to detect beat notes in the region of spectral overlap between SH and SF frequencies (Fig. 5c). We used an InGaAs detector with a longwave cutoff 1.7 μm combined with a long-pass filter (λ > 1.25 μm) to reject the bulk of SH peak. The beat frequency here is related to $f_{ceo}^{OPO}$ through the equation:

$$f_{beat} = 2f_o - (f_o + f_{ceo}^{OPO}) = f_o - f_{ceo}^{OPO} \quad (2)$$

From eq. (1-2) we see that there are two possibilities for the beatnote: $f_{beat} = \frac{1}{2}f_0$ or $f_{beat} = \frac{1}{2}f_0 \pm \frac{1}{2}f_r$. Again, by switching to neighboring OPO cavity-length resonances, we were able to observe beatnotes shifted by $\frac{1}{2}f_r$ from each other (Fig. 5c).

## Conclusion

We achieved a frequency comb spanning from 3 to 12 μm (with 12 μm being the longwave cutoff of our detector) with a smooth spectral distribution, the output power of 245 mW, and conversion efficiency that exceeds 20% from the 2.35-μm driving laser. The key to getting such a broad spectrum was the use of a subharmonic OPO with an unconventional long-wave mode-locked pump and a minimally dispersive cavity (with zero second-order net GDD) through using (i) all-gold-coated mirrors, (ii) a pump injector, and (iii) intracavity elements, including the OP-GaP nonlinear crystal, with zero GVD near the OPO midpoint frequency. Our theoretical analysis and simulations show that hyperparametric processes due to $\chi^{(3)}$ in the OP-GaP substantially contribute to broadening the spectrum. It is important that these processes preserve coherence of the OPO comb, which was confirmed through detecting RF beats in different portions of the spectrum. Finally, a fully stabilized frequency comb based on a Cr:ZnS Kerr-lens mode-locked laser with the capability of referencing to the primary frequency standards has recently been reported [4]. Since a subharmonic OPO is a coherent frequency divider, this opens up the possibility of direct linking the whole ultra-broadband mid-IR comb to the absolute frequency scale. Last but not least, due to a non-dissipative nature of the $\chi^{(2)}$ and $\chi^{(3)}$ processes involved and the total photon recycling, very high, approaching 100%, conversion efficiency from the mode-locked driving oscillator to its subharmonic might be expected in future. This is also supported by the fact that we routinely observe pump depletion in excess of 80%. The ultrabroadband subharmonic OPO system can be used as a powerful new photonic tool in a plethora of metrological and spectroscopic applications.

We acknowledge support from the Office of Naval Research (ONR), grants number N00014-15-1-2659 and N00014-18-1-2176.

## Methods

**Nonlinear single-wave propagation model of the OPO**

To simulate the evolution of the OPO pulses, we used an approach based on the nonlinear wave equation in the frequency domain [64-66]. This equation treats all the interacting waves, including the OPO, the pump, SF, and SH, as a *single* field and describes the dynamics of the electric field rather than that of the envelope. Compared to the conventional approach that uses a set of coupled-wave equations with distinct frequency bands, this method is more suitable to simulate our ultrabroadband subharmonic OPO for the reasons that (i) there are several overlapping spectral bands, e.g. between the pump laser and the OPO, and (ii) the nonlinear crystal can provide simultaneous phase-matching for multiple $\chi^{(2)}$ (including cascading) and $\chi^{(3)}$ (intensity-dependent phase and four-wave mixing) nonlinear processes. Assuming that all the interacting waves are co-polarized, and under plane-wave approximation, the evolution of the circulating electric field in the nonlinear medium is described in the frequency domain as follows [66]:

$$\frac{\partial \tilde{E}(z,\omega)}{\partial z} + ik(\omega)\tilde{E}(z,\omega) = -i\frac{\omega^2}{2\epsilon_0 c^2 k(\omega)} \tilde{P}^{\text{NL}}(z,\omega), \quad (1M)$$

$$\tilde{P}^{\text{NL}}(z,\omega) = \epsilon_0\big(\chi^{(2)}\mathcal{F}\{E(z,t)E(z,t)\} + \chi^{(3)}\mathcal{F}\{E(z,t)E(z,t)E(z,t)\}\big), \quad (2M)$$

where $E(z,t)$ is the time-domain $E$-field, $\tilde{E}(z,\omega)$ is the frequency-domain $E$-field, $\tilde{P}^{\text{NL}}(z,\omega)$ is the nonlinear polarization in the frequency domain, $k(\omega)$ is the wavevector module in the medium, $c$ is the speed of light in vacuum, $\epsilon_0$ is the vacuum dielectric permittivity, $\mathcal{F}\{\}$ represents Fourier transform, and $\chi^{(2)}$ and $\chi^{(3)}$ are quadratic and cubic nonlinear susceptibilities, respectively. The dispersion $\text{Re}[k(\omega)]$ in the nonlinear medium (GaP) is rigorously calculated from its refractive index data [67]. The set of equations (1M-2M) can be numerically solved by the split-step Fourier method [68]. For each propagation step of $\Delta z$, the computing switches between the time and frequency domain to account for nonlinear interaction and linear propagation, respectively. At the exit of the nonlinear crystal ($z = l$), the pulse proceeds to the linear propagation in the cavity introducing only dispersion and loss. The feedback loop is closed after the full roundtrip in the cavity, at the entrance of the nonlinear medium, $z = 0$, where a new pump field is added:

$$\tilde{E}^{(m)}(0,\omega) = \sqrt{R_{inj}(\omega)}\tilde{E}_{\text{in}} + \sqrt{T_{inj}(\omega)}\sqrt{1 - loss}\, e^{i\phi(\omega)}\tilde{E}^{(m-1)}(l,\omega). \quad (3M)$$

Here $m$ is the roundtrip number, $R_{inj}(\omega)$ is the intensity reflection and $T_{inj}(\omega)$ transmission coefficient of the injector, *loss* is the net intensity roundtrip loss (wavelength-independent) in the cavity – from the outcoupling wedge, gold mirrors, scattering, etc., and $\phi(\omega)$ is an extra phase per roundtrip due to the cavity-length mismatch and GDD of the intracavity elements (other than the nonlinear medium). A quantum noise (one photon per mode) was added as an initial input to start the parametric process. For our simulation we assumed that the incoming pump is a bandwidth-limited hyperbolic-secant pulse with a full-width half-maximum (FWHM) duration of 62 fs and zero carrier-envelope offset. The cross-section-averaged peak intensity of the pump inside the nonlinear crystal was set to 50 GW/cm². For the nonlinear refraction $n_2$ of GaP, we took the value of $1.9 \times 10^{-18}$ m²/W, based on $n_2 = 7 \times 10^{-18}$ m²/W measured at λ=1.04 µm [69] scaled to λ≈4.7 µm using the scaling law derived from the two-band model of a semiconductor [70]. Our simulation agrees well with experimental data, including the OPO spectrum, the modified (when the OPO is operating) spectrum of the pump, and the spectrum of the sum frequency (Figs. 3-4). The simulated pump depletion of 79% is close to the experimental of 83%. The model predicts that the average power of the circulating OPO pulse is 1–2 times higher (depending on the outcoupling loss) than that of the incoming pump power. Also, the inset to Fig. 3c shows the simulated time-domain profiles for the electric field and intensity of the OPO waveform suggesting pulse duration of ≈ 20 fs.

## References

1. Schliesser, A., Picqué, N. & Hänsch, T.W. Mid-infrared frequency combs, Nat. Photon. 6, 440-449 (2012).
2. N. Picqué and T. W. Hänsch, Frequency comb spectroscopy, Nat. Photon. 13, 146–157 (2019).
3. K. F. Lee, C. Mohr, J. Jiang, P. G. Schunemann, K. L. Vodopyanov, and M. E. Fermann, Midinfrared frequency comb from self-stable degenerate GaAs optical parametric oscillator, Opt. Express 23, 26596 (2015)
4. S. Vasilyev, V. Smolski, J. Peppers, I. Moskalev, M. Mirov, Y. Barnakov, S. Mirov, and V. Gapontsev, Middle-IR frequency comb based on Cr:ZnS laser Opt. Express. 27, 35079 (2019).
5. T.W. Neely, T.A. Johnson; S.A. Diddams, Mid-infrared optical combs from a compact amplified Er-doped fiber oscillator, Opt. Lett. 36, 4020 (2011).
6. F. Zhu, H. Hundertmark, A. A. Kolomenskii, J. Strohaber, R. Holzwarth, and H. A. Schuessler, High-power midinfrared frequency comb source based on a femtosecond Er:fiber oscillator, Opt. Lett. 38, 2360 (2013).
7. S. A. Meek, A. Poisson, G. Guelachvili, T. W. Hänsch, N. Picqué, Fourier transform spectroscopy around 3 µm with a broad difference frequency comb, Appl. Phys. B 114, 573 (2014).
8. F.C. Cruz, D.L. Maser, T. Johnson, G. Ycas, A. Klose, F.R. Giorgetta, I. Coddington, S.A. Diddams, Mid-infrared optical frequency combs based on difference frequency generation for molecular spectroscopy, Opt. Express 23, 26814 (2015).
9. D. L. Maser, G. Ycas, W. I. Depetri, F. C. Cruz, S. A. Diddams, Coherent frequency combs for spectroscopy across the 3–5 µm region, Appl. Phys. B 123, 142 (2017).
10. K. F. Lee, C. J. Hensley, P. G. Schunemann, and M. E. Fermann, Midinfrared frequency comb by difference frequency of erbium and thulium fiber lasers in orientation-patterned gallium phosphide, Opt. Express 25, 17411 (2017).
11. A. Gambetta, R. Ramponi, and M. Marangoni, Mid-infrared optical combs from a compact amplified Er-doped fiber oscillator, Opt. Lett. 33, 2671 (2008).

12. A. Gambetta, N. Coluccelli, M. Cassinerio, D. Gatti, P. Laporta, G. Galzerano, and M. Marangoni, Milliwattlevel frequency combs in the 8-14 μm range via difference frequency generation from an Er:fiber oscillator, Opt. Lett. 38, 1155 (2013).
13. A. Ruehl, A. Gambetta, I. Hartl, M.E. Fermann, K.S.E. Eikema, and M. Marangoni, Widely-tunable mid-infrared frequency comb source based on difference frequency generation, Opt. Lett. 37, 2232 (2012).
14. F. Keilmann, S. Amarie, Mid-infrared frequency comb spanning an octave based on an Er fiber laser and difference-frequency generation, J. Infrared Milli. Terahz Waves 33, 479 (2012).
15. A. Catanese, J. Rutledge, M. C. Silfies, X. Li, H. Timmers, A. S. Kowligy, A. Lind, S. A. Diddams, and T. K. Allison, Mid-infrared frequency comb with 6.7 W average power based on difference frequency generation, Opt. Lett. 45, 1248 (2020).
16. H. Timmers, A. Kowligy, A. Lind, F. C. Cruz, N. Nader, M. Silfies, G. Ycas, T. K. Allison, P. G. Schunemann, S. B. Papp, and S. A. Diddams, Molecular fingerprinting with bright, broadband infrared frequency combs, Optica 5, 727 (2018).
17. A. S. Kowligy, H. Timmers, A. J. Lind, U. Elu, F. C. Cruz, P. G. Schunemann, J. Biegert, S. A. Diddams, Infrared electric field sampled frequency comb spectroscopy, Sci. Adv. 5, eaaw8794 (2019).
18. I. Pupeza, D. Sánchez, J. Zhang, N. Lilienfein, M. Seidel, N. Karpowicz, T. Paasch-Colberg, I. Znakovskaya, M. Pescher, W. Schweinberger, V. Pervak, E. Fill, O. Pronin, Z.Wei, F. Krausz, A. Apolonski, and J. Biegert, High-power sub-two-cycle mid-infrared pulses at 100 MHz repetition rate, Nat. Photon. 9, 721 (2015).
19. S. Vasilyev, I. S. Moskalev, V. O. Smolski, J. M. Peppers, M. Mirov, A. V. Muraviev, K. Zawilski, P. G. Schunemann, S. B. Mirov, K. L. Vodopyanov, and V. P. Gapontsev, Super-octave longwave mid-infrared coherent transients produced by optical rectification of few-cycle 2.5-μm pulses, Optica 6, 111 (2019).
20. S. Vasilyev, I. Moskalev, V. Smolski, J. Peppers, M. Mirov, A. Muraviev, K. Vodopyanov, S. Mirov, and V. Gapontsev, Multi-octave visible to long-wave IR femtosecond continuum generated in Cr:ZnS-GaSe tandem, Opt. Express 27, 16405 (2019).
21. S. Vasilyev, V.Smolski, I. Moskalev, J. Peppers, M. Mirov, Y. Barnakov, V. Fedorov, D. Martyshkin, A. Muraviev, K. Zawilski, P. Schunemann, S. Mirov, K. Vodopyanov, V. Gapontsev, Multi-octave infrared femtosecond continuum generation in Cr:ZnS-GaSe and Cr:ZnS-ZGP tandems, Proc. SPIE 11264, 1126407 (2020).
22. Q. Wang, J. Zhang, A. Kessel, N. Nagl, V. Pervak, O. Pronin, and K. F. Mak, Broadband mid-infrared coverage (2–17 μm) with few-cycle pulses via cascaded parametric processes, Opt. Lett. 44, 2566 (2019).
23. J. Zhang, K. Fai Mak, N. Nagl, M. Seidel, D. Bauer, D. Sutter, V. Pervak, F. Krausz, and O. Pronin, Multi-mW, few-cycle mid-infrared continuum spanning from 500 to 2250 cm$^{-1}$, Light Sci. Appl. **7**, 17180 (2018).
24. J. Zhang, K. Fritsch, Q. Wang, F. Krausz, K. F. Mak, and O. Pronin, "Intra-pulse difference-frequency generation of mid-infrared (2.7–20 μm) by random quasi-phase-matching," Optics Letters 44, 2986-2989 (2019).
25. A. J. Lind, A. Kowligy, H. Timmers, F. C. Cruz, N. Nader, M. C. Silfies, T. K. Allison, and S. A. Diddams, Mid-infrared frequency comb generation and spectroscopy with few-cycle pulses and $\chi^{(2)}$ nonlinear optics, Phys. Rev. Lett. 124, 133904 (2020).
26. F. Adler, K. C. Cossel, M. J. Thorpe, I. Hartl, M. E. Fermann, and J. Ye, Phase-stabilized, 1.5 W frequency comb at 2.8-4.8 μm, Opt. Lett. 34, 1330 (2009).
27. K. Iwakuni, G. Porat, T. Q. Bui, B. J. Bjork, S. B. Schoun, O. H. Heckl, M. E. Fermann, J. Ye, Phase-stabilized 100 mW frequency comb near 10 μm, Applied Physics B 124, 128 (2018).
28. L. Maidment, O. Kara, P. G. Schunemann, J. Piper, K. McEwan, D. T. Reid, Long-wave infrared generation from femtosecond and picosecond optical parametric oscillators based on orientation-patterned gallium phosphide, Appl. Phys. B 124, 143 (2018).
29. N. Granzow, M. A. Schmidt,W. Chang,L. Wang, Q. Coulombier, J. Troles, P. Toupin, I. Hartl, K. F. Lee, M. E. Fermann, L. Wondraczek and P. St.J. Russell, Mid-infrared supercontinuum generation in $As_2S_3$-silica nano-spike step-index waveguide, Opt. Express 21, 10969 (2013).
30. B. Kuyken, T. Ideguchi, S. Holzner, M. Yan, T. W. Hänsch, J. Van Campenhout, P. Verheyen, S. Coen, F. Leo, R. Baets, G. Roelkens, and N. Picqué, An octave-spanning mid-infrared frequency comb generated in a silicon nanophotonic wire waveguide, Nature Commun. 6, 6310 (2015).
31. H. Guo, C. Herkommer, A. Billat, D. Grassani, C. Zhang, M. H. P. Pfeiffer, W. Weng., C.-S. Brès and T. J. Kippenberg, Mid-infrared frequency comb via coherent dispersive wave generation in silicon nitride nanophotonic waveguides, Nat. Photon. 12, 330 (2018).
32. N. Nader, A. Kowligy, J. Chiles, E. J. Stanton, H. Timmers, A. J. Lind, F. C. Cruz, D. M. B. Lesko, K. A. Briggman, S. W. Nam, S. A. Diddams, and R. P. Mirin, Infrared frequency comb generation and spectroscopy with suspended silicon nanophotonic waveguides, Optica 6, 1269 (2019).
33. N. Nagl, K. F. Mak, Q. Wang, V. Pervak, F. Krausz, and O. Pronin, Efficient femtosecond mid-infrared generation based on a Cr:ZnS oscillator and step-index fluoride fibers, Opt. Lett. 44, 2390 (2019).
34. D. Martyshkin, V. Fedorov, T. Kesterson, S. Vasilyev, H. Guo, J. Liu, W. Weng, K. Vodopyanov, T. J. Kippenberg, S. Mirov, Visible-near-middle infrared spanning supercontinuum generation in a silicon nitride ($Si_3N_4$) waveguide, Opt. Mat. Express 9, 2553 (2019).
35. A. Hugi, G. Villares, S. Blaser, H. C. Liu, and J. Faist, Midinfrared frequency comb based on a quantum cascade laser, Nature 492, 229 (2012).
36. Q. Lu, D. Wu, S. Slivken, and M. Razeghi, High efficiency quantum cascade laser frequency comb, Sci. Rep. 7, 43806 (2017).
37. M. Bagheri, C. Frez, L. A. Sterczewski, I. Gruidin, M. Fradet, I. Vurgaftman, C. L. Canedy, W. W. Bewley, C. D. Merritt, C. S. Kim, M. Kim, and J. R. Meyer, Passively mode-locked interband cascade optical frequency combs, Sci. Reports 8, 3322 (2018).
38. A. G. Griffith, R. K. W. Lau, J. Cardenas, Y. Okawachi, A. Mohanty, R. Fain, Y. H. D. Lee, M. Yu, C. T. Phare, C. B. Poitras, A. L. Gaeta, M. Lipson, Silicon-chip mid-infrared frequency comb generation, Nature Commun. 6, 6299 (2015).
39. A. G. Griffith, M. Yu, Y. Okawachi, J. Cardenas, A. Mohanty, A. L. Gaeta, and M. Lipson, Coherent mid-infrared frequency combs in silicon-microresonators in the presence of Raman effects, Opt. Express. 24, 13044 (2016).
40. C.Y. Wang, T. Herr, P. Del'Haye, A. Schliesser, J. Hofer, R. Holzwarth, T.W. Hänsch, N. Picqué, T.J. Kippenberg, Mid-infrared optical frequency combs at 2.5 µm based on crystalline microresonators, Nature Commun. 4, 1345 (2013).
41. C. Lecaplain, C. Javerzac-Galy, E. Lucas, J. D. Jost, and T. J. Kippenberg, Quantum cascade laser Kerr frequency comb, ArXiv:1506.00626, 2015
42. A. A. Savchenkov, V. S. Ilchenko, F. Di Teodoro, P. M. Belden, W. T. Lotshaw, A. B. Matsko, and L. Maleki, Generation of Kerr combs centered at 4.5 µm in crystalline microresonators pumped with quantum-cascade lasers, Opt. Lett. 40, 3468 (2015)
43. M. Yan, P.-L. Luo, K. Iwakuni, G. Millot, T. W. Hänsch, and N. Picqué, Mid-infrared dual-comb spectroscopy with electro-optic modulators, Light: Science & Applications 6, e17076 (2017).
44. K. L. Vodopyanov, S. T. Wong, and R. L. Byer, "Infrared frequency comb methods, arrangements and applications", U.S. patent 8,384,990 (February 26, 2013).
45. N. Leindecker, A. Marandi, R.L. Byer, K. L. Vodopyanov, K. L. Broadband degenerate OPO for mid-infrared frequency comb generation. Opt. Express 19, 6304–6310 (2011).
46. N. Leindecker, A. Marandi, R.L. Byer, K. L. Vodopyanov, J. Jiang, I. Hartl, M. Fermann, and P. G. Schunemann, Octave-spanning ultrafast OPO with 2.6-6.1 µm instantaneous bandwidth pumped by femtosecond Tm-fiber laser, Opt. Express 20, 7047-7053 (2012).
47. Q. Ru, Z. E. Loparo, X. Zhang, S. Crystal, S. Vasu, P. G. Schunemann, and K. L. Vodopyanov, Self-referenced octave-wide subharmonic GaP optical parametric oscillator centered at 3 µm and pumped by an Er-fiber laser, Opt. Lett. 42, 4756-4759 (2017).
48. Q. Ru, K. Zhong, N. P. Lee, Z. E. Loparo, P. G. Schunemann, S.Vasiyev, S. B. Mirov, K. L.

Vodopyanov, Instantaneous spectral span of 2.85 - 8.40 μm achieved in a Cr:ZnS laser pumped subharmonic OPO, Proc. SPIE 10088, 1008809 (2017).

49. E. Sorokin, A. Marandi, P. G. Schunemann, M. M. Fejer, R. L. Byer, and I. T. Sorokina, Efficient half-harmonic generation of three-optical-cycle mid-IR frequency comb around 4 μm using OP-GaP, Opt. Express 26, 9963 (2018).

50. V. Smolski, S. Vasilyev, I. Moskalev, M. Mirov, Q. Ru, A. Muraviev, P. Schunemann, S. Mirov, V. Gapontsev, and K. Vodopyanov, Half-Watt average power femtosecond source spanning 3–8 μm based on subharmonic generation in GaAs, Appl. Phys. B 124, 101 (2018).

51. A. V. Muraviev, V. O. Smolski, Z. E. Loparo, and K. L. Vodopyanov, "Massively parallel sensing of trace molecules and their isotopologues with broadband subharmonic mid-infrared frequency combs", Nature Photon. 12, 209 (2018).

52. Z. E. Loparo, E. Ninnemann, Q. Ru, K. L. Vodopyanov, and S. S. Vasu, Broadband mid-infrared optical parametric oscillator for dynamic high-temperature multi-species measurements in reacting systems, Opt. Lett. 45, 491-494 (2020)

53. A. Marandi, N. C. Leindecker, K. L. Vodopyanov, and R. L. Byer, All-optical quantum random bit generation from intrinsically binary phase of parametric oscillators, Opt. Express 20, 19322–19330 (2012).

54. A. Marandi, Z. Wang, K. Takata, R. L. Byer, and Y. Yamamoto, Network of time-multiplexed optical parametric oscillators as a coherent Ising machine, Nature Photon. 8, 937–942 (2014).

55. M. Jankowski, A. Marandi, C. R. Phillips, R. Hamerly, Kirk A. Ingold, R. L. Byer, and M. M. Fejer, Temporal simultons in optical parametric oscillators, Phys. Rev. Lett. 120, 053904 (2018).

56. Marandi, A., Leindecker, N., Pervak, V., Byer, R.L. and Vodopyanov, K.L. Coherence properties of a broadband femtosecond mid-IR optical parametric oscillator operating at degeneracy, Opt. Express 20, 7255–7262 (2012).

57. K. F. Lee, N. Granzow, M. A. Schmidt, W. Chang, L. Wang, Q. Coulombier, J. Troles, N. Leindecker, K. L. Vodopyanov, P. G. Schunemann, M. E. Fermann, P. St. J. Russell, and I. Hartl, Midinfrared frequency combs from coherent supercontinuum in chalcogenide and optical parametric oscillation, Opt. Lett. 39, 2056-2059 (2014).

58. V. O. Smolski, H. Yang, S. D. Gorelov, P.G. Schunemann, and K.L. Vodopyanov, Coherence properties of a 2.6–7.5 μm frequency comb produced as a subharmonic of a Tm-fiber laser, Opt. Lett. 41, 1388 (2016).

59. C. Wan, P. Li, A. Ruehl, and I. Hartl, Coherent frequency division with a degenerate synchronously pumped optical parametric oscillator, Opt. Lett. 43, 1059 (2018).

60. A. Marandi, K. A. Ingold, M. Jankowski, and R. L. Byer, Cascaded half-harmonic generation of femtosecond frequency combs in the mid-infrared, Optica 3, 324–327 (2016).

61. K. L. Vodopyanov, Laser-based Mid-infrared Sources and Applications (Wiley, 2020).

62. W. C. Hurlbut and Y.-S. Lee, K. L. Vodopyanov, P. S. Kuo, and M. M. Fejer, Multiphoton absorption and nonlinear refraction of GaAs in the mid-infrared, Opt. Lett. 32, 668 (2007).

63. O. H. Heckl, B. J. Bjork, G. Winkler, P. B. Changala, B. Spaun, G. Porat, T. Q. Bui, K. F. Lee, J. Jiang, M. E. Fermann, P. G. Schunemann, and J. Ye, Three-photon absorption in optical parametric oscillators based on OP-GaAs, Opt. Lett. 41, 5405-5408 (2016).

64. T. Brabec and F. Krausz, Nonlinear optical pulse propagation in the single-cycle regime, Phys. Rev. Lett. 78, 3282 (1997).

65. M. Conforti, F. Baronio, and C. De Angelis, Modeling of ultrabroadband and single-cycle phenomena in anisotropic quadratic crystals, J. Opt. Soc. Am. B 28, 1231 (2011).

66. H. Guo, X. Zeng, and M. Bache, Generalized nonlinear wave equation in frequency domain, arXiv:1301.1473 (2013).

67. J. Wei, J. M. Murray, J. O. Barnes, D. M. Krein, P. G. Schunemann, and S. Guha, Temperature dependent Sellmeier equation for the refractive index of GaP, Opt. Mater. Express 8, 485 (2018).

68. G. Agrawal, Nonlinear Fiber Optics, Optics and Photonics (Elsevier Science, 2012).

69. F. Liu, Y. Li, Q. Xing, L. Chai, M. Hu, C. Wang, Y. Deng, Q. Sun, and C. Wang, Three-photon absorption and Kerr nonlinearity in undoped bulk

GaP excited by a femtosecond laser at 1040 nm, J. Opt. **12**, 095201 (2010).

70. M. Sheik-Bahae, D. C. Hutchings, D. J. Hagan, and E. W. Van Stryland, Dispersion of bound electron nonlinear refraction in solids, IEEE J. Quantum Electron. **27**, 1296 (1991).